\documentclass[11pt,twoside]{utarticle}
\usepackage{amssymb}
\usepackage{floatflt,graphicx} 
\usepackage{verbatim,epsfig}
\numberwithin{equation}{section}

\newcommand{\abs}[1]{\lvert #1\rvert}

\newdimen\tableauside\tableauside=1.0ex
\newdimen\tableaurule\tableaurule=0.4pt
\newdimen\tableaustep
\def\phantomhrule#1{\hbox{\vbox to0pt{\hrule height\tableaurule width#1\vss}}}
\def\phantomvrule#1{\vbox{\hbox to0pt{\vrule width\tableaurule height#1\hss}}}
\def\sqr{\vbox{%
  \phantomhrule\tableaustep
  \hbox{\phantomvrule\tableaustep\kern\tableaustep\phantomvrule\tableaustep}%
  \hbox{\vbox{\phantomhrule\tableauside}\kern-\tableaurule}}}
\def\squares#1{\hbox{\count0=#1\noindent\loop\sqr
  \advance\count0 by-1 \ifnum\count0>0\repeat}}
\def\tableau#1{\vcenter{\offinterlineskip
  \tableaustep=\tableauside\advance\tableaustep by-\tableaurule
  \kern\normallineskip\hbox
    {\kern\normallineskip\vbox
      {\gettableau#1 0 }%
     \kern\normallineskip\kern\tableaurule}%
  \kern\normallineskip\kern\tableaurule}}
\def\gettableau#1 {\ifnum#1=0\let\next=\null\else
  \squares{#1}\let\next=\gettableau\fi\next}

\begin{document}

\thispagestyle{empty}
\leftline{\copyright~ 1998 International Press}
\leftline{Adv. Theor. Math. Phys. {\bf 2} (1998) {\ 1405 --1439 } }
\begin{center}
\vspace{0.2in}
{\huge \bf Non-Supersymmetric Conformal 
\\[5mm] Field Theories from Stable \\[5.mm] Anti-de Sitter Spaces}
\vspace{0.2in}
\renewcommand{\thefootnote}{}
\footnotetext{\small e-print archive: {\texttt
http://xxx.lanl.gov/abs/hep-th//9810206}}
\renewcommand{\thefootnote}{}
\footnotetext{Work supported in part by NSF Grant PHY9511632 and
  the Robert A.~Welch}\footnotetext{Foundation.}

{\bf Jacques Distler and Frederic Zamora}
  \vspace{0.2in}

Theory Group, Physics Department\\
    University of Texas at Austin\\
    Austin TX 78712 USA.\\
{\tt distler@golem.ph.utexas.edu}\\
{\tt zamora@zerbina.ph.utexas.edu}

\end{center}
\begin{abstract}
We describe new non-supersymmetric conformal field theories in
three and four dimensions, using the CFT/AdS correspondence.
In order to believe in their existence at large $N_c$ and strong
't Hooft coupling, we explicitly check the stability of the
corresponding non-supersymmetric anti-de Sitter backgrounds.
Cases of particular interest are the relevant deformations of the
${\cal N}=4$ SCFT in $SU(3)$ and $SO(5)$ invariant directions. It
turns out that the  former is a stable, and the latter an {\it unstable}
non-supersymmetric type IIB background.
\end{abstract}

\newpage
\pagenumbering{arabic}
\setcounter{page}{1406}

\pagestyle{myheadings}
\markboth{\it NON-SUPERSYMMETRIC CONFORMAL FIELD ...}{\it J. DISTLER, F. ZAMORA}
\section{Introduction}

Two dimensional conformal field theories exhibit a very rich structure, and
much can be said about their properties. One of the most striking features is
the ability to control, not just the conformal theory itself, but also the
nearby \emph{nonconformal} theories obtained by adding certain specially-chosen
relevant perturbations. These nonconformal theories exhibit a structure
(exact-integrability) almost as rich as that of their conformal cousins.
Moreover, one can actually follow these perturbations and see how the theory
flows from the ultraviolet fixed point CFT to a
new fixed point (a different conformal field theory) in the infrared.

Not so much is known about conformal field theories in higher dimensions. The
richest information, to date, has only come for supersymmetric conformal field
theories.

Recently, however, a new avenue to understanding conformal field theories (in
various dimensions) has opened up. The CFT/AdS correspondence
\cite{Maldacena:1997re}  gives a precise quantitative relation
\cite{Witten:AdSholography,Gubser:GaugeCorrelators}
between a conformal field theory in $d$ dimensions, and a solution to string
theory/M-theory on a background of the form $AdS_{d+1}\times M$, with $M$
compact. At least for large $N$ (and large $g_sN$),
the latter is computable,
in that it is well-approximated by a classical supergravity calculation.

In this paper, we would like to imitate the construction which was very
successful in two dimensions, and construct new conformal field theories by
taking known examples from the CFT/AdS correspondence, perturbing them by
relevant operators, and flowing to the infrared. In favourable circumstances,
we will find that the perturbed theory indeed flows to a new conformal field
theory in the infrared.

In our examples, the perturbation breaks some or all of the
supersymmetry. Hence, we will, in particular, be learning about new,
\emph{non-supersymmetric} conformal field theories.
This is exciting, in itself,
as there are not too many known examples of such for $d>3$. We will find
examples of these non-supersymmetric CFTs in $d=3,4$ and discuss the case
of $d=6$.

The downside of breaking supersymmetry is that self-consistency of the
supergravity solution is no longer assured, even at the classical level.
If a tachyonic scalar in anti-de Sitter
space has a negative mass-squared
which exceeds the
Freedman-Breitenlohner stability bound
\cite{Breitenlohner:GaugedSUGRAstability,Mezincescu:1985ev,Mezincescu:1984iu},
leads to an expo-nentially-growing mode, rendering the supergravity solution
unstable. Super-symmetry ensures that there are no such unstable modes  in a
supersymmetry-preserving solution to the supergravity equations. It,
plausibly, also takes care that quantum corrections do not upset the
classical solution. (We say, \emph{plausibly}, as no one actually knows how to
compute quantum corrections in these Ramond-Ramond backgrounds.) In the
non-supersymmetric case, we will actually have to check the stability, even at
the classical level \emph{by hand}. We will also restrict ourselves to the
large $N$ limit, where quantum corrections are parametrically small
(being down
by factors of $1/N^2$) and do not upset the large $N$ solution.

We will study in greatest detail the
relevant deformations of the $d=4$ ${\cal N}=4$ SCFT.
Those relevant operators are mapped to the scalars
in supergravity multiplet, whose dynamics is encoded in the
five dimensional ${\cal N}=8$ $SU(4)$ gauged supergravity
Lagrangian \cite{Gunaydin:Gauged5dSUGRA}.
In particular, in \S\ref{sec:SU(3)}, we follow the renormalization
group flow of the ${\cal N}=4$ SYM with the addition of a
mass term for one of the gluinos. Such a term breaks all
the supersymmetries and the $SU(4)_R$ R-symmetry to
$SU(3)$. We will see the existence of a non-supersymmetric
$SU(3)$ invariant background $AdS_5 \times {\tilde S}^5$
of type IIB string theory, smoothly connected
(via the VEV of the scalar field in the supergravity corresponding to the
gaugino mass) with the  usual BPS $SU(4)$ invariant $AdS_5 \times S^5$
background. We will see the relation between the supergravity solution which
interpolates between these asymptotic behaviours and RG flow in the field
theory, perturbed by this  $SU(3)$-invariant relevant operator.
We compute the mass spectrum of the low-lying states in this $SU(3)$-invariant
supergravity solution (equivalently, we find the conformal dimension of the
corresponding operators in the $d=4$ CFT), and check that it is, indeed, a
stable solution.

 In \S\ref{sec:SO(5)} we analyse another kind of relevant deformation
of the ${\cal N}=4$ SYM: a quadratic term for one of the
adjoint scalars. This term also breaks supersymmetry, and
leaves an $SO(5)$ subgroup of $SU(4)_R$ unbroken.
As before, its ultraviolet relevant coupling constant connects the
maximally supersymmetric $AdS_5 \times S^5$ type IIB background
to a non-supersymmetric $AdS_5 \times {\tilde S}^5$
background (with now ${\tilde S}^5$ being an squashed five-sphere
in one direction, breaking the isometries $SO(6)$ to $SO(5)$).
Unfortunately, this supergravity background proves to be unstable. So, in this
case, there is no corresponding $SO(5)$-invariant $d=4$ CFT.

\S\ref{sec:3D} deals with $d=3$ conformal field theories.
We review some $SU(3)$ invariant fixed points studied by Warner
\cite{Warner:NewExtremaScalarPot}, and  observe the existence of a
non-supersymmetric
$SO(3)\times SO(3)$ invariant fixed point, connected to the ${\cal
N}=8$ $SO(8)$ invariant
fixed point by an appropriate renormalization group flow.

 Finally, in \S\ref{sec:6D}, we discuss the relevant
deformation of the $d=6$ ${\cal N}=(2,0)$ SCFT,
breaking supersymmetry and the $SO(5)$ R-symmetry
group to $SO(4)$. As in the $d=4$ $SO(5)$ invariant
theory, this deformation ends up in an unstable
supergravity background.
We also check the stability of a
non-supersymmetric M-theory background
\cite{Berkooz:1998qp}, which was recently proposed as leading to a
non-supersymmetric,
$SO(4)$ invariant, CFT in $d=6$. This theory, too, turn out to be unstable.

 After this work was completed, we received the preprint
\cite{Girardello:1998pd}, which overlaps with material of
\S\ref{sec:SU(3)},\ref{sec:SO(5)}. They also noted the existence of the
$SU(3)$ and
$SO(5)$ invariant  type IIB backgrounds and their connection to conformal
field theories on the boundary. Unfortunately, they did not compute the
spectrum
of masses around these new critical points and so could not check the
stability of
these solutions.

\section{Relevant Deformations of
\boldmath{$D=4$, ${\cal N}=4$} SCFT}

The CFT/AdS correspondence can be extended to particular
non-super-symmetric relevant directions of the ${\cal N}=4$
SCFT. At large $N_c$ and strong 't Hooft
coupling, the deformed theory is given by the
solution of the classical equations of motion of the
${\cal N}=8$ $SU(4)$ gauged supergravity action in $5D$,
with boundary conditions determined by the relevant couplings.

There are two type of relevant deformations: by $\Delta =2$
superconformal primaries
\begin{equation}
\lambda_{IJ} \int dx^4 \ ({\rm tr} (X^I X^J) - \frac{\delta^{IJ}}{6}
{\rm tr}\boldsymbol{X}^2) \,,
\end{equation}
with $X^I$ being the six real scalars in the adjoint of $SU(N_c)$
and in the ${\bf 6}$ of $SU(4)$;
and by $\Delta=3$ superconformal primaries
\begin{equation}
m_{AB} \int dx^4 \ {\rm tr}(\lambda^A \lambda^B) \ + {\rm h.c.} \,,
\end{equation}
with $\lambda^A$ being Weyl spinors in the adjoint of
$SU(N_c)$ and in the ${\bf 4}$ of $SU(4)$.
These deformations  give, respectively,
supersymmetry-breaking masses to the scalars and to the gluinos.

The $\lambda_{IJ}$ and $m_{AB}$ are in the ${\bf 20}'$
and ${\bf 10_c}$ of $SU(4)$, respectively. These,
plus the two $SU(4)$
singlet marginal couplings tr$(F^2)$ and tr$(F\wedge F)$,
are associated to the $42$ scalars
of the ${\cal N}=8$ supergravity multiplet in $D=5$.
To study the field theory  deformed by these relevant couplings, we need to
study the dynamics
of the supergravity theory with these scalars turned on.

The ungauged ${\cal N}=8$, $D=5$ supergravity Lagrangian
has global symmetry $E_{6(6)}$ and local symmetry $Sp(4)$, where $Sp(4)$ is
the maximal compact subgroup of $E_{6(6)}$.
The previous 42 scalars are described by an element
${\cal U}^{\alpha\beta}{}_{ab}$ of the coset space $E_{6(6)}/Sp(4)$
which transforms in the ${\bf 27}$ of $E_6$ (acting on the
$E_6$ indices $\alpha,\beta=1,\dots,8$ from the left) and ${\bf 27}$ of
$Sp(4)$ (acting on the $Sp(4)$ indices $a,b=1,\dots,8$ from the right)
\footnote{The ${\bf 27}$ ($\tableau{1 1}$) of $Sp(4)$ is the traceless,
antisymmetric tensor representation,
$Z^{ab}= -Z^{ba}$, $\Omega_{ab}Z^{ab}=0$.}
After gauging the $SU(4)$ subgroup of $E_{6(6)}$,
a nontrivial scalar potential is generated.
It is proportional to the square of the $SU(4)$ gauge coupling,
$g$, and breaks $E_{6(6)}$ down to $SU(4)\times SL(2,\BR)$
\cite{Gunaydin:Gauged5dSUGRA}.
We will use an $SU(4)\times U(1)\subset Sp(4)$ subgroup of $E_{6(6)}$ as a
basis to represent the
42 scalars into ${\cal U}^{ab}{}_{cd}$ (the $E_{6(6)}$ indices are
restricted to be $Sp(4)$ indices $a,b,c,d=1,\dots,8$)
\footnote{The reference \cite{Gunaydin:Gauged5dSUGRA}
chose the $SL(6,\BR)\times SL(2,\BR)$ subgroup of $E_6$
as the basis to perform the gauging of $SU(4)\simeq SO(6)$.
As we will see, the complex basis described here is more convenient for
studying
of the $\Delta=3$ deformations;
and the real $SL(6,\BR)\times SL(2,\BR)$ basis
is more suited for the $\Delta=2$ deformations.}.
In the $SU(4)$ unitary gauge,
only the $42$ physical scalars remain in the
$27\times 27$ matrix ${\cal \Phi}$, with
${\cal U}= {\rm exp}({\cal \Phi})$.
The scalars transform in the irreducible ${\bf 42}$, $\tableau{1 1 1 1}$,
of $Sp(4)$
The ${\bf 42}$ acts on the ${\bf 27}$ of $Sp(4)$ by
$Z^{ab} \rightarrow {\cal U}^{ab}{}_{cd}\ Z^{cd}$.
The $SU(4)$ gauge group and the $U(1)_{\chi}$ compact generator of
$SL(2,\BR)$ are embedded in this $Sp(4)$,
with the following branching rules:
\begin{equation}\label{eq:su4reps}
\begin{split}
{\bf 8} &= {\bf 4}_{1} + \overline{\bf 4}_{-1}
\\
{\bf 27} &= {\bf 15}_{0} + {\bf 6}_{2} + {\bf 6}_{-2}
\\
{\bf 42} &= {\bf 20}_{0}' + {\bf 10}_{-2} + \overline{\bf 10}_{2}
+ {\bf 1}_{4} + {\bf 1}_{-4} \,,
\end{split}
\end{equation}
where the subscript denotes the $U(1)_{\chi}$ charge.
Ordering the $27$ dimensional vector space by
$Z^{ab}=
\left(
\begin{smallmatrix}
{\bf 15}_{0}\\
{\bf 6}_{2}\\
{\bf 6}_{-2}
\end{smallmatrix}
\right)
$, the
${\bf 42}$ can be written in block form
\begin{equation}
\Phi =
\begin{pmatrix}
{\bf 20}_{0}' & {\bf 10}_{-2} & \overline{\bf 10}_{2} \\
\overline{\bf 10}_{2} & {\bf 20}_{0}' & {\bf 1}_{4} \\
{\bf 10}_{-2} & {\bf 1}_{-4} & {\bf 20}_{0}' \\
\end{pmatrix}
\end{equation}
where the ${\bf 20}'$ is the representation $\tableau{2 2}$ of $SU(4)$.

The supergravity scalar potential is built from the
quantities \cite{Gunaydin:Gauged5dSUGRA}:
\begin{subequations}
\begin{align}
{\cal W}_{abcd} &= {\cal P}_{efgh} \,
{\cal U}^{ef}{}_{ab}\, {\cal U}^{gh}{}_{cd}
\label{eq:Wabcd}
\\
{\cal W}_{ab} &= {\cal W}_{acbd} \, \Omega^{cd} \,,
\end{align}
\end{subequations}
where $\Omega^{cd}$ is the $8\times 8$ symplectic metric and
${\cal P}_{efgh}$ is the $SU(4)\times U(1)_\chi$-invariant, skew-symmetric,
 bilinear quadratic form
$\left(
\begin{smallmatrix}
0&0&0\\
0&0&-\Bid\\
0&\Bid&0
\end{smallmatrix}
\right)$ acting on\newline

$\left(
\begin{smallmatrix}
{\bf 15}_{0}\\
{\bf 6}_{2}\\
{\bf 6}_{-2}
\end{smallmatrix}
\right)
$.
The final expression for the potential is
\footnote{$Sp(4)$ indices are raised and lowered with the symplectic metric
$\Omega^{ab}$.}
\begin{equation}\label{eq:genpot}
V(\Phi) = \frac{g^2}{16}\left(
\frac{1}{2}{\cal W}^{abcd} {\cal W}_{abcd} -
{\cal W}^{ab} {\cal W}_{ab} \right) \,.
\end{equation}
Then, the scalar plus graviton sector of the gauged supergravity
Euclidean Lagrangian (in appropriate $5D$ Planck units)
is described by the following
non-linear sigma model coupled to gravity:
\begin{equation}\label{eq:sugra}
{\cal L} = {\sqrt g} \left\{ -\tfrac{1 }{ 4} R
- \tfrac{1}{24}{\rm tr}
\left((\nabla_{\mu}{\cal U}^{-1})(\nabla^{\mu}{\cal U})\right) + V(\Phi)
\right\} \,.
\end{equation}

\section{The Non-Supersymmetric $\boldsymbol{SU(3)}$-Invariant
Theory}\label{sec:SU(3)}

As a first step in exploring  the non-supersymmetric
theories connected to the ${\cal N}=4$ fixed point, one can study
the theory in the $SU(3)$-invariant direction given by the
relevant deformation:
\begin{equation}\label{eq:fermion}
S_{\rm rel} = m \int d^4x \ {\rm tr}(\lambda^4\lambda^4) \ + {\rm h.c.}
\end{equation}
The relevant coupling $m$ has UV scaling mass dimension one
and it gives a mass term only to the gaugino $\lambda^4$.
Supersymmetry is completely broken and the $SU(4)$ R-symmetry
is broken to a global $SU(3)$ symmetry. It is useful to keep track of the
$U(1)$ factor, a linear combination of the $U(1)_m \subset SU(4)$
and the $U(1)_\chi$, which is unbroken by this perturbation (but which is
broken by the VEV of the dilaton).

To study the theory along this $SU(3)$ invariant direction, we decompose the
$SU(4)$ representations discussed above under
$SU(3) \times U(1)_m\subset SU(4)$:
\begin{equation}
\begin{split}
{\bf 4} &= {\bf 3}_{1} + {\bf 1}_{-3}
\\
{\bf 6} &= {\bf 3}_{-2} + \overline{\bf 3}_{2}
\\
{\bf 10} &= {\bf 6}_{2} + {\bf 3}_{-2} + {\bf 1}_{-6}
\\
\overline{\bf 10} &= {\bf 6}_{-2}+ \overline{\bf 3}_{2} + {\bf 1}_{6}
\\
{\bf 15} &= {\bf 8}_{0} + {\bf 3}_{4} + \overline{\bf 3}_{-4}
+ {\bf 1}_{0}
\\
{\bf 20}' &= {\bf 8}_{0} + {\bf 6}_{-4} + \overline{\bf 6}_{4} \,.
\end{split}
\end{equation}
Together with \eqref{eq:su4reps}, we can identify all the operators
in the different $SU(3) \times U(1)_m \times U(1)_\chi$ representations.
They can also be derived from their microscopic field content.
The four gluinos $\lambda^A$ decompose as $\lambda^i \in {\bf 3}_{(1,-1)}
$ ($i=1,2,3$) and $\lambda^4 \in {\bf 1}_{(-3,-1)}$.
The six scalars $X^I$ can be combined into complex combinations
$Z^i, {\overline Z_j}$, which transform as the
${\bf 3}_{(-2,0)}+\overline{\bf 3}_{(2,0)}$.
The relevant perturbation
${\cal O}_{44} = {\rm tr}(\lambda^4 \lambda^4)$ is in the
$SU(3)$ singlet representation ${\bf 1}_{(-6,-2)}$,
so that it is left invariant by the
$U(1)$ factor generated by the charge $Q= Q_m -3Q_\chi$.

Now we turn to the supergravity description of this
non-supersymmetric $SU(3)$ invariant theory.
Only the two complex scalars corresponding to the
$SU(3)$ singlet representations have boundary conditions
different from the trivial ones, as the only couplings which are turned on
on the field theory side are $m$, $g_{YM}$
and $\theta_{YM}$. The  (complex) ${\bf 1}_{(0,4)}$
field, corresponding to the non-compact generators of $SL(2,\BR)$
(the dilaton and the axion), do not appear in the potential $V(\Phi)$
because of the $SL(2,\BR)$ invariance of the ${\cal W}_{abcd}$
tensor \eqref{eq:Wabcd}. We write this complex field as a modulus times a
phase, $\rho\ex{i\alpha}$.
The other complex scalar,  $\sigma \ex{i\varphi}$, is associated to the complex
coupling $m$.
The phase is eaten via the Higgs mechanism, and, in a unitary gauge,
only the modulus enters in the coset element
${\cal U}$. We order the $SU(3)\times U(1)_m \times U(1)_\chi$
representations  in the ${\bf 27}$ by
\begin{equation*}
\begin{pmatrix}
{\bf 1}_{(0,0)}\\
{\bf 3}_{(4,0)}\\
\overline{\bf 3}_{(-4,0)}\\
{\bf 8}_{(0,0)}\\
{\bf 3}_{(-2,2)}\\
\overline{\bf 3}_{(2,2)}\\
{\bf 3}_{(-2,-2)}\\
\overline{\bf 3}_{(2,-2)}
\end{pmatrix}\,.
\end{equation*}
In this basis, it proves convenient to parametrize $\CU$ as
\begin{equation}\label{eq:u0}
{\cal U}_0(\sigma,\rho,\alpha) =
\ex{X_0}\ex{X_1}
\end{equation}
where,\vfill 
\begin{eqnarray}
X_0&=&\begin{pmatrix}
0 & 0 & 0 & 0 & 0 & 0 & 0 & 0 \\
0 & 0 & 0 & 0 & 0 & 0 & 0 & 0 \\
0 & 0 & 0 & 0 & 0 & 0 & 0 & 0 \\
0 & 0 & 0 & 0 & 0 & 0 & 0 & 0 \\
0& 0& 0& 0& 0& 0& \rho\ex{i\alpha}& 0 \\
0 & 0 & 0 & 0 & 0 & 0 & 0 & \rho\ex{i\alpha} \\
0 & 0 & 0 & 0 & \rho\ex{-i\alpha} & 0 & 0 & 0 \\
0& 0& 0& 0& 0& \rho\ex{-i\alpha}& 0& 0
\end{pmatrix},\nonumber \\
X_1&=&\begin{pmatrix}
0 & 0 & 0 & 0 & 0 & 0 & 0 & 0 \\
0 & 0 & 0 & 0 & 0 & 0 & \sigma & 0 \\
0 & 0 & 0 & 0 & 0 & \sigma & 0 & 0 \\
0 & 0 & 0 & 0 & 0 & 0 & 0 & 0 \\
0 & 0 & 0 & 0 & 0 & 0 & 0 & 0 \\
0 & 0 & \sigma & 0 & 0 & 0 & 0 & 0 \\
0 & \sigma & 0 & 0 & 0 & 0 & 0 & 0 \\
0 & 0 & 0 & 0 & 0 & 0 & 0 & 0
\end{pmatrix} ,\,.
\end{eqnarray}

Plugging this into the supergravity Lagrangian \eqref{eq:sugra},
one gets
\begin{equation}
{\cal L} = \sqrt{g} \left( -\frac{R}{ 4} + \tfrac{1}{ 2}(\partial \sigma)^2
+\tfrac{1}{ 2}(\partial \rho)^2+\tfrac{1}{ 2}\sinh^2(\rho)(\partial
\alpha)^2  +V(\sigma) \right)
\end{equation}
where $V(\sigma) = \frac{3}{8 r^2} W(\sigma)$, and
\begin{equation}\label{eq:pote}
W(\sigma) = {\rm cosh}^2(2\sigma) -4\, {\rm cosh}(2\sigma) -5 \,.
\end{equation}
The classical $SU(4)$ gauge coupling has been determined to be
$g^2 =4/3r^2$ in order to have anti-de Sitter radius
$r^2 = (4\pi g^2_{YM} N_c)^{1/2} \alpha'$ at the
supersymmetric point $\sigma=0$.

Note that the metric for $\rho$ and $\alpha$ is $SL(2,\BR)$ invariant, as it
should be. Weak string coupling corresponds to $\rho\to\infty$.

As already mentioned, the phase of $\sigma$ is eaten by the Higgs mechanism.
It is not a modulus.  On the field theory side, this corresponds to the
fact that  the phase of $m$ can be
removed by a chiral rotation of the gluinos ({\it i.e.}~
absorbed by a shift in the
$\theta_{YM}$ angle).

\subsection{The equations of motion}\label{sec:eom}

Now, we just have to solve the equations of motion for
$\{g_{mn}, \sigma, \rho,\alpha\}$ subject to the boundary conditions as $z
\to 0$:
\begin{equation}\label{eq:bound}
\begin{split}
g_{mn} &\to \frac{r^2 }{ z^2} \delta_{mn}
\\
\sigma &\to mz
\\
\rho&\to \rho_0\\
\alpha&\to \alpha_0 \,.
\end{split}
\end{equation}
\begin{figure}
\centerline{\psfig{file=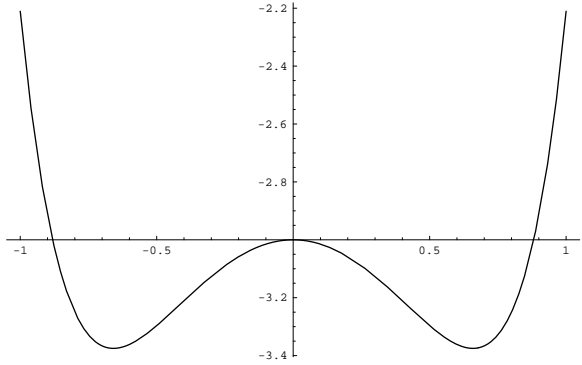,width=3.0in}}
\caption{The scalar potential $V_{SU(3)}(\sigma)$ in $r$
units.}\label{fig:su3}
\end{figure}
The boundary conditions are translationally invariant along
the four dimensional Euclidean space $x^\mu$. Therefore,
we consider the ansatz
\begin{equation}\label{eq:me1}
\begin{split}
\sigma &= \sigma(z, m)
\\
ds^2 &= \frac{r^2}{z^2} ( e^{g(z,m)} dz^2 + e^{h(z,m)} d{\bf x}^2 ) \,,
\end{split}
\end{equation}
with $\sigma = g = h = 0$ for $m=0$.
In fact, the equations of motion and the boundary conditions
only depend of the dimensionless
combination $t \equiv mz$, therefore the field solutions only
depend on this parameter $t$.
Furthermore, one can show that the equations of motion
do not admit a zero of $e^g$
(which would yield a singularity of the metric).
Hence we can redefine the coordinate $z$ such that $g=0$
and the boundary conditions \eqref{eq:bound} are still satisfied.
Finally one ends up with the equations (prime means derivative with
respect  to $t$):
\begin{subequations}
\begin{align}
\sigma'' + \left(2 h' -\frac{3}{t}\right)\sigma'
- \frac{3}{8t^2}\, \frac{dW(\sigma)}{d\sigma} &=0
\label{eq:T}
\\
h'' + \frac{h'}{t} + \frac{4}{3}{\sigma'}^2 &=0
\label{eq:E2}
\\
{h'}^2 -4\frac{h'}{t} +\frac{4}{t^2}
-\frac{2}{3}{\sigma'}^2 +\frac{1}{2 t^2} W(\sigma) &=0\\
\rho''+\left(2h'-\frac{3}{t}\right)\rho'&=0\\
\alpha'' +\left(2h'-\frac{3}{t}+2\rho'\coth\rho\right)\alpha'&=0 \,.
\label{eq:E1}
\end{align}
\end{subequations}

The dilaton and axion equations are homogeneous, and are easily solved by
taking $\rho=\mathrm{const}=\rho_0$, $\alpha=\mathrm{const}=\alpha_0$. There
remain three equations for two unknown functions, $h(t)$ and $\sigma(t)$.
But one can prove that one of them is a consequence
of the others two, and the system admits a solution.

Solving the equations of motion \eqref{eq:T}--\eqref{eq:E1}
for any $t$ gives the behavior of the deformed field theory
for any value of $m$. In general, the spacetime metric $g_{mn}(z,m)$
will not be anti-de Sitter, and the boundary theory is not conformally
invariant
for finite $m$.

But there is an interesting phenomenon in
the asymptotic behavior $t \to \infty$ of the
spacetime metric.
This behavior is quite easy to obtain from the equations of motion.
For $t \to \infty$, the solution becomes
\begin{equation}\label{eq:asy}
\begin{split}
h(t) &\simeq (2 -b) {\rm ln}(t) + \frac{h_{-2c} }{ t^{2c}} + \dots
\\
\sigma &\simeq \sigma_0 + \frac{\sigma_{-c} }{t^c} + \dots
\end{split}
\end{equation}
where $\sigma_0 = -\frac{1}{2}{\rm arcosh}(2)$, $c=b(\sqrt{3}-1)$ and
\begin{equation}\label{eq:b}
\frac{b^2}{4} =
\frac{V(\sigma_0)}{ V(0)} = \frac{9}{8} >1.
\end{equation}
This result requires some explanation.

\begin{figure}

\centerline{\psfig{file=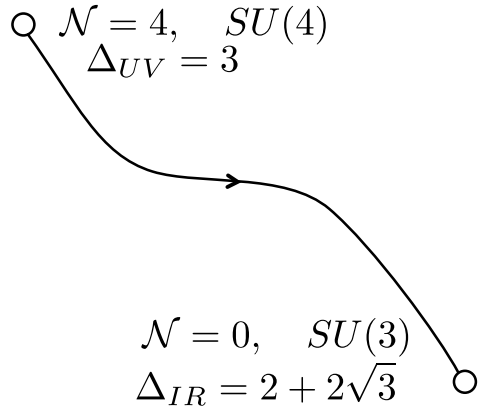,width=2.5in}}
\caption{The RG flow along the $SU(3)$ invariant
direction. $\Delta$ is the conformal dimension of the perturbation at the
fixed points.}\label{fig:rg}
\end{figure}
The scalar potential $V(\sigma)$ has two critical points
where $dV/d\sigma=0$ (see fig.~\ref{fig:su3}).
One is at $\sigma=0$ and it is a maximum of the potential.
It gives an $AdS_5$ space, with radius $r$,
coming from the compactification of type IIB on $S^5$,
which preserves all the supersymmetries of type IIB.
On the 4D field theory side, it
corresponds to the $SU(4)$ invariant ${\cal N}=4$ superconformal
field theory.

The other critical point is located
at ${\cosh}(2\sigma_0) =2$ and it is where
the solution for $t\to \infty$ ends up.
One can see that for $\sigma \not=0$, the theory
does not have invariant Killing spinors \cite{Gunaydin:Gauged5dSUGRA},
so supersymmetry is completely broken.
We can look at the infinitesimal
$t\to 0$ solution \eqref{eq:bound}
and the asymptotic $t\to \infty$ solution \eqref{eq:asy}
as the solution driven for $m\to 0$ and $m\to \infty$,
respectively, and general coordinate $z$. Keeping the external
probe energies fixed, these limits correspond
to the ultraviolet and infrared limits, respectively.
Then, the $SU(3)$ critical point at $\sigma=\sigma_0$
is associated to an infrared fixed point of the ${\cal N}=4$ SYM
deformed theory by the $SU(3)$-invariant relevant deformation
\eqref{eq:fermion}.

By performing the coordinate transformation
\begin{equation}\label{eq:coord}
\begin{split}
t' &= t^{b/2}
\\
x' &= \frac{2}{ b}x
\end{split}
\end{equation}
the asymptotic solution of the metric \label{asym} becomes
\begin{equation}
ds^2 = \frac{r_{IR}^2 }{ {z'}^2}({dz'}^2 + \sum_{i=1}^4 {dx'_i}^2)
\end{equation}
with
\begin{equation}\label{eq:rads}
r_{IR}=\frac{2}{b} \,r \,;
\end{equation}
i.e., anti-de Sitter space, but with a smaller radius
than the one in the ultraviolet, $r_{UV}=r$.
The proportionality factor  is
given by the ratio of the two corresponding cosmological constants
\eqref{eq:b}.
As the bulk theory is again (Euclidean) anti-de Sitter, just with smaller
radius, we have $SO(1,5)$ isometry for that metric, yielding conformal
invariance for the boundary field theory in the infrared.

The $SU(3)$ critical point
corresponds to a type IIB background  AdS$_5 \times M_5$,
with $M_5$ a stretched five-sphere \cite{Romans:IIBnewComp}, and the
anti-de Sitter radius being $r_{IR} = \frac{ 2}{ b}r$.
This background breaks all the supersymmetries and
the compact manifold $M_5$ has the isometry group $SU(3)$.

\subsection{The Wilson loop}

 For general $m \not=0$, we have a deformed metric of the type
\begin{equation}\label{eq:metric}
ds^2 = \frac{{\hat r}^2 \alpha' }{ z^2}
(dz^2 + e^{h(z,m)} \sum_{i=1}^4 dx_i^2) \,.
\end{equation}

Now consider a Nambu-Goto string action propagating in that
deformed 5D space,
with the world-sheet boundary attached at $z=0$.
This computes the expectation value
of a Wilson loop in the boundary theory
\cite{Maldacena:AdSWilsonLoops,Rey:1998ik}. We will
consider the static and symmetric Wilson loop
\begin{gather*}
x_1= x_2= {\rm const}
\\
x_3 =\sigma \, , \quad x_4 = \tau
\\
z= z(\sigma) \,,
\end{gather*}
with
\begin{gather*}
z(0)= z(L) = 0
\\
z(L - \sigma) = z( \sigma)
\\
z(\frac{L }{ 2}) = z_M \, , \quad {\dot z}(\frac{L}{2}) = 0 \,;
\end{gather*}
such that $z_M$ is the maximum distance reached by the string.

This gives the quark and anti-quark energy potential
\begin{equation}\label{eq:action}
E(L) = \frac{r^2 }{ \pi} \int_0^{L/2} d\sigma
\ \frac{e^{h/2}}{z^2}\sqrt{(e^h + {\dot z}^2)}
\end{equation}
evaluated through the geodesic. Observe that we are dealing with
a classical mechanics problem: the movement of a particle in one
dimension with initial point at $(\tau_i, z_i) = (0, 0)$
and end point at $(\tau_f, z_f) = (L/2, z_M)$.
The conserved ``particle energy'' is
\begin{equation}\label{eq:energy}
{\cal E} = -\frac{{\hat r}^2 e^{3h/2} }{ \pi z^2 \sqrt{e^h +{\dot z}^2}} =
- \frac{{\hat r}^2 e^{h_M}}{ \pi z_M^2} \,,
\end{equation}
where $h_M = h(z_M)$.

If we move the end point
$\delta\tau_f = \delta L/2$, $\delta z_f = \delta z_M$, the
quark and anti-quark energy changes as
\begin{equation}\label{eq:force}
\frac{\delta E }{ \delta L} = \frac{{\hat r}^2 e^{h_M} }{ 2\pi z_M^2} \,.
\end{equation}
So, the force between the quark and the anti-quark is given by the
``particle energy''. We want to read this force for $L \to \infty$.
In order do it, we need the relation between $L$ and $z_M$.
{}From \eqref{eq:energy} one gets
\begin{equation}
\frac{L}{ 2} = z_M \int_0^1 dy \, \frac{y^2 e^{-h/2} }{
\sqrt{e^{2(h-h_1)} - y^4}} \,,
\end{equation}
where $y=z/z_M$.

The major contribution to the integral
is at $y \sim 1$. One can evaluate the integral in that region,
taking $z_M \to \infty$, such that
$h(y)\simeq (2-b)( \ln(m z_M) + \ln y)$. The result is
\begin{equation}
\frac{L}{ 2} \simeq e^{-h_M/2} z_M I \,,
\end{equation}
with
\begin{equation}\label{eq:Int}
I = \int_0^1 dy \ \frac{y^{3b/2 -1} }{ \sqrt{1 - y^{2b}}} \,.
\end{equation}
This gives the relation, for $m L \to \infty$,
\begin{equation}\label{eq:scaling}
m L \sim (m z_M)^{b/2} \,.
\end{equation}

For $m L >> 1$, the force between electric sources becomes
\begin{equation}
\frac{\delta E }{ \delta L} \simeq \frac{m^2}{(m z_M)^b}
\sim \frac{1}{ L^2} \,.
\end{equation}
It is a nontrivial result that
the asymptotic behavior of the deformed metric gives
a Coulomb potential. This is another indication that the infrared theory is
conformally-invariant. However, at this stage, it might well turn out
to be a free theory. One indication that it is not free is that the Peskin
exponent
\footnote{At a second order phase transition for a gauge theory in
$d$ dimensions, $E(L)=2\pi(d-4+\eta)/L$ \cite{Peskin:1980ay}.}
$\eta$ at the $SU(3)$ theory:
\begin{equation}
\eta_{SU(3)} = \frac{I[b=3/\sqrt{2}]}{I[b=2]}(2 +\eta_{SU(4)}) -2
\simeq 0.9428 \, \eta_{SU(4)} -0.1144 \,,
\end{equation}
with $I[b]$ given by the integral (\ref{eq:Int}), does not vanish, as one
expects for a free theory.

We can make additional checks of the Coulomb behavior.
We should also see the same kind of
potential for magnetic sources. We can test that, computing
the minimal area expanded by a D1 brane attached to the boundary
$z=0$, which gives a 't Hooft loop. The D1 action is
\begin{equation}
S_{D1} = \frac{1}{ 2\pi \alpha'}\int d^2\sigma \, e^{-\Phi}
\sqrt{{\rm det}(g_{ij}\partial_\alpha X^i \partial_\beta X^j)}
\end{equation}
where $\Phi$ is the dilaton.
It gives the same force between magnetic sources, up to
an additional dilaton dependent factor,
\begin{equation}
\frac{\delta E_m }{ \delta L}=
\frac{{\hat r}^2 e^{h_M-\Phi_M} }{ 2\pi z_M^2} \,.
\end{equation}
For $h$ and $\sigma$ being the unique solution of
\eqref{eq:T}--\eqref{eq:E1},
with the asymptotic behavior \eqref{eq:asy}, $\Phi(z) =\Phi_0$
is an exact solution of the equations of motion with the appropriate boundary
conditions.
Then, the force between two monopoles becomes
\begin{equation}
\frac{\delta E_m }{ \delta L}=
\frac{(4\pi g_{YM}^2 e^{-2\Phi_0})^{1/ 2} e^{h_M} }{ 2\pi z_M^2} \,.
\end{equation}

Taking into account that $g_{YM}^2=e^{\Phi_0}$, one gets exactly
the same force as the one between electric sources, but with the Yang-Mills
coupling inverted: $g_{YM} \to 1/g_{YM}$. This result is consistent with
the electric-magnetic duality of the Coulomb phase.

As an additional check, we can compute the
normalized solutions of the dilaton equation for
$\Phi = f(z)e^{ikx}$. The eigenvalues $-k^2$
give the masses of the intermediate states in the two
point correlation function of the operator
${\rm tr}(F^2)$ \cite{Witten:AdSfiniteTempQCD}. The equation is
\begin{equation}\label{eq:dil}
f'' +\left(2h' -\frac{3}{ z}\right)f' -k^2 e^{-h} f =0 \,.
\end{equation}

For $z \simeq 0$, one chooses the normalizable solution $f \sim z^4$.
For $z \to \infty$, the normalizable asymptotic behavior is
$f \sim 1/z^a$, with $a>0$. The equation \eqref{eq:dil} gives
\begin{equation}
\frac{a(a+ 2b) }{ z^{a+2}} -k^2\frac{m^{b-2}}{z^{a +2 -b}} =0
\end{equation}
with unique solution for $k^2=0$ and $a=0$.
As the smooth deformation by $m$ does not
change the topology of the five-dimensional spacetime,
there are no non-zero square-integrable
solutions in the deformed metric.

\subsection{The nontrivial fixed point
and the stability of supergravity solution}

The Coulombic behaviour of the Wilson loop that we found in the previous
subsection is indicative of the conformal invariance of the infrared
theory. But
it is compatible with the infrared theory being a {\it free}
conformal theory. More exciting would be a nontrivial interacting conformal
theory. In the presence of supersymmetry, a wealth of such four
dimensional conformal field theories are known, {\it e.g.}~the
Argyres-Douglas points of
${\cal N}=2$  QCD \cite{Argyres:N=2FixedPointsInD=4}, the ${\cal N}=1$ QCD in
the conformal window $3N_c/2 < N_f < 3N_c$ \cite{Seiberg:EMDuality} and, of
course, the ${\cal N}=4$ SYM,
both at the origin of the moduli space.  More recently, non-supersymmetric
conformal field theories  have been proposed as orbifolds of these in the
CFT/AdS point of view \cite{Kachru:1998ys}. We will presently see that the
solution we found in the previous subsections is a  nontrivial interacting
fixed point without supersymmetry.

 At $m=0$ we are at the ${\cal N}=4$ fixed point. As we increase
$m$, we go away from that fixed point. For
$m \to \infty$ we will end up on a new fixed point.
Sending $m \to \infty$ can be seen as going to the infrared,
as we saw by introducing the dimensionless variable $t=m z$.

In \S\ref{sec:eom} we derived, for $t\to \infty$
\begin{subequations}
\begin{align}
ds^2 &= (m r)^2 \left(\frac{1 }{ t^2}dz^2
+ \frac{1 }{ t^b}\sum_{i=1}^4 dx_i^2 \right)
\label{eq:asymetric}
\\
\sigma &\sim \sigma_0
\label{eq:asytac}
\\
m L &\sim t^{b/2}\label{eq:rel} \,.
\end{align}
\end{subequations}
There are two ways to look at this:
\begin{itemize}
\item[1)] for $m$ fixed, it gives
the infrared behavior
$z\to \infty$ (that sends $L\to \infty$);
\item[2)] for $m \to \infty$, it is the solution of
the equations of motion for any $z\not=0$ (and also for
any $L\not=0$ in the field theory side).
\end{itemize}
Taking the second point of view, it is better to work
in terms of the coordinates $\{z',{\bf x}'\}$
defined in (\ref{eq:coord}).
The 4D field theory lives at the boundary $z'=0$.
Following the standard procedure as in \cite{Witten:AdSholography},
one gets that for an scalar field $\phi$ with mass $m_\phi$ at the
$SU(3)$ point, its behavior for $z'\to 0$ is
\begin{equation}
\phi \to {z'}^{-\lambda_+} \phi_0 \,,
\end{equation}
with $\lambda_+$ the larger root of the equation
\begin{equation}\label{eq:quad}
\lambda(\lambda +4) = m_\phi^2 r^2_{\rm IR} \,.
\end{equation}
The boundary field value $\phi_0$ couples to a conformal
operator which has the scaling mass dimension
\begin{equation}
\Delta = 4 + \lambda_+= 2 + \sqrt{4 + m_\phi^2 r^2_{IR}} \,.
\end{equation}

In general this gives anomalous dimensions,
a sure sign that the fixed point is nontrivial.
If $b$ were equal to $2$,
we would get the same formulae as in the ${\cal N}=4$
fixed point. But $b=\frac{3}{\sqrt{2}}$ for this $SU(3)$ invariant fixed
point. The scaling dimensions of the operators will be different than  they
were in the $SU(4)$ supersymmetric point -- first because the
masses at the new
extremum of the supergravity potential are different, and second because $b$ is
different. We have already seen that the mass of $\sigma$ has changed. At the
supersymmetric point, it was tachyonic (the corresponding operator in the
4-dimensional SCFT is relevant). At the $SU(3)$ point, $\sigma$ has positive
mass-squared, and the corresponding operator is irrelevant
(see fig.~\ref{fig:rg}).

In fact, we will commonly obtain irrational anomalous
dimensions (see tables~\ref{tab:scalars} and \ref{tab:rest} below). As we are
dealing with a non-supersymmetric CFT, the erstwhile chiral primaries
(whose scaling dimensions were protected in the supersymmetric theory, and
equal to their free field values)  are no longer protected.
Observe also that we need $m^2 \geq -b^2/r^2$ to have
real solutions for \eqref{eq:quad}. This is also the bound for vacuum
stability in
a five-dimensional AdS background with radius $r_{IR} = 2r/b$
\cite{Witten:AdSholography,Breitenlohner:GaugedSUGRAstability}. An unstable
supergravity solution does not correspond to a sensible CFT on the boundary,
as the scaling dimensions corresponding to the unstable modes are complex.
So, both from the point of view of checking the stability of the
non-supersymmetric solution to the 5D supergravity equations, and to extract
the physics of the purported new fixed point,
we need to compute the complete  quadratic
dependence of the supergravity potential in all the  scalars fields.

For the parametrization of the coset space $E_{6(6)}/Sp(4)$,
it is convenient to choose one that gives canonical kinetic terms.
We take

\begin{equation}\label{eq:param}
{\cal U} = {\cal U}_0(\sigma,\rho,\alpha) \, \ex{X} \,,
\end{equation}

with ${\cal U}_0(\sigma,\rho,\alpha)$ given by \eqref{eq:u0}.
The $27\times 27$ hermitian matrix $X$ contains the fluctuations
of the remaining tachyonic (at $\sigma=0$) scalars.
We parametrize them by the following fields:
a hermitian and traceless $3\times 3$ matrix $h^i_j$ for the
${\bf 8}_{(0,0)}$ representation; symmetric and complex $3\times 3$
matrices $t^{ij}$ and $s^{ij}$
for the ${\bf 6}_{(-4,0)}$ and
${\bf 6}_{(2,-2)}$ representations,
respectively; and a complex three vector $v^i$ for the
${\bf 3}_{(-2,-2)}$. The vectors $v^i$, are eaten
by the Higgs mechanism (along with the phase of $\sigma$).
So, in a unitary gauge, they are absent from the
scalar potential. The rest of the fields appear in $X$ as
 \begin{equation}\label{eq:x6}
X =
\begin{pmatrix}
0 & 0 & 0 & 0  & 0 & 0 & 0
& 0 \\ 0 & h & 0 & {\sqrt 2}\,{\overline t^\curlyvee} & 0
& {\sqrt 6}\,s & 0 & 0 \\
0 & 0 & {\overline h} & {\sqrt 2}\,t^\curlyvee & 0
& 0 & {\sqrt 6}\,{\overline s} & 0 \\
0 & {\sqrt 2}\,t^\curlyvee
& {\sqrt 2}\,{\overline t^\curlyvee}
& h_{\rm adj} & 0 & 0 & 0 & 0 \\
0 & 0 & 0 & 0 & h & {\sqrt 2}\,t & 0 & 0 \\
0 & {\sqrt 6}\,{\overline s}
& 0 & 0
& {\sqrt 2}\,{\overline t} & {\overline h} & 0 & 0 \\
0 & 0 & {\sqrt 6}\,s
& 0 & 0 & 0 & h & {\sqrt 2}\,t \\
0 & 0 & 0 & 0 & 0 & 0 & {\sqrt 2}\,{\overline t} & {\overline h}
\end{pmatrix}
\end{equation}
where the bar means complex conjugation and
$(t^\curlyvee)^{ij}_k = \epsilon^{ijl} \, {\overline t}_{lk}$.

The broken $SU(4)$ generators give rise to massive $W$ bosons in the
${\bf 3}+\overline{{\bf 3}}+{\bf 1}$ of $SU(3)$.
The masses are easily computed as a function of $\sigma$:
\begin{equation}
\begin{align}
m^2_{\bf 3} &= \frac{2}{9r^2}(\cosh(\sigma) - 1)
\\
m^2_{\bf 1}& = \frac{1}{6r^2}(\cosh(2 \sigma) - 1) \,.
\end{align}
\end{equation}
The aforementioned 7 scalars from the ${\bf 42}$ become the longitudinal
components of these massive $W$ bosons.

With the parametrization
\eqref{eq:param}
\footnote{We work to quadratic order in the fluctuations
of $h$, $s$ and $t$, but to all orders in $\sigma$, $\rho$ and $\alpha$.},
the kinetic terms for the remaining scalars become:
\begin{equation}
\begin{split}
-\frac{1}{24}\mathrm{tr}\bigl((\partial U^{-1})\partial U \bigr) =&
\frac{1}{2} (\partial \sigma)^2 +\frac{1}{2} (\partial \rho)^2
+ \frac{1}{2}\sinh^2(\rho) (\partial \alpha)^2
\\
&+ \mathrm{tr}\bigl((\partial {\overline s})(\partial s)\bigr)
+ \mathrm{tr}\bigl((\partial {\overline t})(\partial t)\bigr)
+\frac{1}{2}\mathrm{tr}\bigl((\partial h)^2\bigr) \,.
\end{split}
\end{equation}
Then, we can read directly the masses from the quadratic terms
in the potential. We have ($p \equiv {\rm cosh}(2\sigma)$):
\begin{equation}
\begin{split}
V^{\rm (quadr)} = \frac{1}{r^2}\Bigl( & \frac{3}{8}(p^2 -4 p -5)
- 3\,\mathrm{tr}({\overline s}s )
\\
&+\frac{1}{4}(7 p^2 -10 p -13)\,\mathrm{tr}({\overline t} t)
+ \frac{1}{3}(4p^2 -5p -5)\,\mathrm{tr}(h^2 ) \Bigr) \,.
\end{split}
\end{equation}

We observe that all the fields satisfy the stability bound
$m_\phi^2 \geq -4/r_{IR}^2 = -9/2 r^2$ of the $SU(3)$
critical point $p=2$
(see table~\ref{tab:scalars} for their corresponding conformal dimensions).
So the non-supersymmetric type IIB
background $AdS_5 \times M_5$ is a stable solution
(at least at large $N_c$ and $g^2_{YM}N_c$).
It defines a non-supersymmetric interacting infrared
fixed point of the $SU(3)$ invariant theory.

To close this discussion of the $SU(3)$ infrared fixed point,
in table~\ref{tab:rest} we give the conformal dimensions
of the scaling operators corresponding
to the remaining modes in the 5D supergravity multiplet
(their mass terms are given in \cite{Gunaydin:Gauged5dSUGRA}).
The graviton is mapped to the energy-momentum tensor $T_{\mu\nu}$ of the
$SU(3)$ invariant CFT. The eight gravitini (in the ${\bf 4}_1$ of
$SU(4)\times U(1)_\chi$) map to the
four complex supersymmetry currents $S_\mu^{A}$ .
The twelve `self-dual' 2-forms (in the ${\bf 6}_{-2}$)
correspond to the operators \cite{Ferrara:AdSandSYM}
\begin{equation}
B^{AB}_{\mu\nu} = \lambda^A \sigma_{\mu\nu}
\lambda^B + 2\,i\,{\rm tr}(Z^{AB} F^+_{\mu\nu}) \,.
\end{equation}
The $SU(4)$ global currents are $J^A_{\mu\,B}$. At the $SU(3)$
point, the currents $J^i_{\mu\,4}$ and $J^4_{\mu\,4}$ are not
conserved, and pick up anomalous dimensions.

\begin{table}[ht]
\begin{center}
\begin{tabular}{|c|c|c|c|}
\hline
CFT operator&Supergravity field&$SU(3)\times
U(1)$&$\Delta_{UV}$\\   
\hline\hline
$\abs{\mathrm{tr}(\lambda^4 \lambda^4)} $&$\sigma$&
${\bf 1}_{0}$  & $3$\\   
$\mathrm{tr}(\lambda^i \lambda^j) $&$s^{ij}$&
${\bf 6}_{8}$  & $3$ \\ 
\hline
$\mathrm{tr}(Z^i Z^j) $&$t^{ij}$&
${\bf 6}_{-4}$  & $2$\\  
$\mathrm{tr}(
Z^{i}\overline{Z}_{j}-\tfrac{\delta^{i}{}_{j}}{3}
\overline{\boldsymbol{Z}}\boldsymbol{Z})
$&$h^i{}_j$&
${\bf 8}_{0}$  & $2$    \\
\hline
\end{tabular}

\vspace{0.1in}

\begin{tabular}{|c|c|}
\hline
CFT operator&$ \Delta_{IR}$\\
\hline\hline
$\abs{\mathrm{tr}(\lambda^4 \lambda^4)} $
&   $2 + 2\sqrt{3} = 5.4641 \cdots $\\
$\mathrm{tr}(\lambda^i \lambda^j) $& $2 + 2/\sqrt{3} = 3.1547 \cdots $\\
\hline
$\mathrm{tr}(Z^i Z^j)$ 
& $2 + \sqrt{26}/3 = 3.6997 \cdots $\\
$\mathrm{tr}(
Z^{i}\overline{Z}_{j}-\tfrac{\delta^{i}{}_{j}}{3}
\overline{\boldsymbol{Z}}\boldsymbol{Z})
$& $2 + \tfrac{2}{3}\sqrt{31/3} = 4.1430 \cdots $\\
\hline
\end{tabular}
\end{center}
\caption{Conformal dimensions of the spinless scaling operators corresponding
to the physical scalars in the 5D supergravity multiplet
at the $SU(3)$ point.}\label{tab:scalars}
\end{table}
 
Finally, there are the spin-$1/2$ scaling operators:
\begin{equation}
\begin{split}
\chi^A &= \sigma^{\mu\nu}{\rm tr}(F^-_{\mu\nu}\lambda^A)
\\
\chi^{AB}_{\quad C} &= \frac{1}{2}\epsilon^{ABDE}
(\overline{Z}_{DE}\overline{\lambda}_C
+ \overline{Z}_{CE}\overline{\lambda}_D )
\end{split}
\end{equation}
in the ${\bf 4}_{-3}$ and ${\bf 20}_{1}$ of $SU(4)\times U(1)_\chi$,
respectively.
These correspond to the 48 symplectic Majorana spinors
of the five-dimensional supergravity multiplet.
At the non-supersymmetric point, eight of them are
eaten by the gravitini. This phenomenon is manifested in the
$SU(3)$ CFT by the anomalous dimensions of the broken supersymmetry
currents $S^A_\mu$ (see table 2)

\section{The $\boldsymbol{SO(5)}$ Invariant Theory}\label{sec:SO(5)}

Having succeeded in finding a new nontrivial critical point when the
$\Delta=3$ relevant perturbation is turned on, we now try turning on,
instead, the $\Delta=2$
relevant perturbation,
\begin{equation}\label{eq:pertso5}
S_{\rm rel} = \mu^2 \int d^4 x \,
\bigl( {\rm tr}(X^6X^6)-\tfrac{1 }{ 6}{\rm tr}\boldsymbol{X}^2 \bigr) \,.
\end{equation}
This gives a positive mass-squared to $X^6$, and formal negative mass-squared
to the scalars
$X^1,
\dots,X^5$.
Supersymmetry is completely broken and $SU(4)$  is broken to $SO(5)$.

Decomposing the scalars in the supergravity under $SO(5)\subset SU(4)$,
\begin{equation}
{\bf 20}'={\bf 1}+{\bf 5}+{\bf 14} \,,
\end{equation}
and the ${\bf 10}$ and $\overline{\bf 10}$ of $SU(4)$ both become the ${\bf
10}$ ($\tableau{1 1}$) of $SO(5)$. The perturbation we are describing
corresponds to turning on the
$SO(5)$ singlet scalar, which we will call $\psi$, in the ${\bf 20}'$.

Since all of the representations of $SO(5)$ that we encounter are {\it real},
it behooves us to choose a real basis in which to parametrize the coset matrix
${\cal U}\in E_{6(6)}/Sp(4)$. The ${\bf 27}$ of $E_6$ decomposes as $({\bf
15},{\bf 1})+({\bf 6},{\bf2})$ under $SO(6)\times SL(2,\BR)$. In a real basis,
the compact $U(1)_\chi\subset SL(2,\BR)$ is generated by
$\left(\begin{smallmatrix}0&-1\\ 1&0\end{smallmatrix}\right)$, and a general
matrix in the coset $SL(2,\BR)/U(1)$ takes the form
\begin{equation}
T=\ex{Y},\qquad Y=\rho\begin{pmatrix}\cos\alpha&\sin\alpha\\
\sin\alpha&-\cos\alpha \end{pmatrix} \,.
\end{equation}

The ${\bf 20}'$ of $SO(6)$ is represented by a
traceless symmetric matrix, $X^I{}_J$. Let
$S=\ex{X}$. If we do not also turn on any of the scalars
in the ${\bf 10}$, then
the $E_{6(6)}$ coset element is block diagonal,
\begin{equation}
{\cal U}=\begin{pmatrix}{\cal U}^{IJ}{}_{KL}&0\\
0&{\cal U}^{I\alpha}{}_{J\beta}
\end{pmatrix} \,,
\end{equation}
where
\begin{equation*}
\begin{split}
{\cal U}^{IJ}{}_{KL} &= 2 S^{[I}{}_{[K} S^{J]}{}_{L]}
\\
{\cal U}^{I \alpha}{}_{J \beta} &= S^I{}_J T^{\alpha}{}_{\beta} \,.
\end{split}
\end{equation*}

Using the general expression \eqref{eq:genpot}, the potential reads
\begin{equation}\label{eq:20pot}
V(S) = - \frac{g^2}{32}\left( ({\rm tr}M)^2 - 2 {\rm tr}(M^2) \right)
\end{equation}
with the symmetric matrix $M= S S^T$.
Observe that $T$ does not appear in the potential, as expected.

\begin{figure}

\centerline{\psfig{file=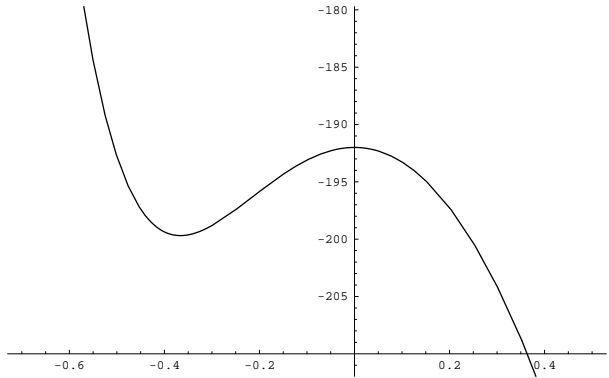,width=3.0in}}
\caption{The scalar potential $V_{SO(5)}(\psi)$ in $r$ units.}
\label{fig:v}
\end{figure}

The $SO(5)$ singlet scalar associated to the relevant perturbation
\eqref{eq:pertso5} enters in $S$ as
\begin{equation*}
S = {\rm diag}(e^{\psi/\sqrt{15}}, \dots, e^{\psi/\sqrt{15}},
e^{-5\psi/\sqrt{15}}) \,.
\end{equation*}
The normalization is chosen to have mass $m_\psi^2= -4/r^2$
at the supersymmetric point $\psi=0$.
The Lagrangian \eqref{eq:sugra} simplifies to:
\begin{equation}\label{eq:Lagr}
\begin{split}
{\cal L} &= \sqrt{g} \bigl\{ -\tfrac{1 }{ 4} R + \tfrac{1}{2}(\partial
\rho)^2 +
\tfrac{1}{2}\sinh^2(\rho)(\partial\alpha)^2
+ \tfrac{1}{2}(\partial \psi)^2 -\frac{1}{8 r^2} W(\psi) \bigr\}
\\
W(\psi) &= 15 e^{4 \psi /\sqrt{15}}
+ 10 e^{-8 \psi/\sqrt{15}} - e^{-20 \psi/\sqrt{15}} \,.
\end{split}
\end{equation}
Now, as in the $SU(3)$ case, we have to solve the equations of motion of
$\{g_{mn},\psi,\rho,\alpha\}$ subject to the boundary conditions for
$z \to 0$:
\begin{equation}\label{eq:asympSO(5)}
\begin{split}
g_{mn} &\to \frac{r^2 }{ z^2} \delta_{mn}
\\
\rho &\to \rho_0
\\
\alpha&\to\alpha_0
\\
\psi &\to - \mu^2 z^2 \,.
\end{split}
\end{equation}
The sign convention in \eqref{eq:asympSO(5)} is to end up at the $SO(5)$
critical point of the scalar potential (see fig.~\ref{fig:v}).

\newpage
\begin{table}[ht]
\begin{center}
\begin{tabular}{|c|c|c|c|}
\hline
CFT operator & Supergravity field & $SU(3)\times
U(1)$&$\Delta_{UV}$\\ 
\hline\hline
$T_{\mu\nu}$ & Graviton & ${\bf 1}_0$ & 4\\
\hline
$S^{i}_\mu$ & Gravitini & ${\bf 3}_{-2}$ & 7/2 \\
$S^{4}_\mu$ & Gravitino & ${\bf 1}_{-6}$ & 7/2 \\
\hline
$B^{i4}_{\mu\nu}$ & 2-forms & ${\bf 3}_{-8}$ & 3 \\
$B^{ij}_{\mu\nu}$ & 2-forms & ${\bf 3}_{-4}$ & 3 \\
\hline
$J^i_{\mu\,j}$ & Gauge bosons & ${\bf 8}_0$ & $3$\\
$J^i_{\mu\,4}$ & W bosons & ${\bf 3}_{4}$ & $3$ \\
$J^4_{\mu\,4}$ & W boson & ${\bf 1}_0$ & $3$\\
\hline
$\chi^i$ & Spinors & ${\bf 3}_{10}$ & $7/2$ \\
\hline
$\chi^{i\bar{j}}_{\ 4}$ & Spinors & ${\bf 8}_6$ & $5/2$\\
$\chi^{ij}_{\ \bar{k}}$ & Spinors & ${\bf 6}_{-2}$ & $5/2$
\\
$\chi^{i4}_{\ \bar{4}}$ & Spinors & ${\bf 3}_{-2}$
& $5/2$ \\
\hline
\end{tabular}


\begin{tabular}{|c|c|}
\hline
CFT operator &$ \Delta_{IR}$ \\
\hline\hline
$T_{\mu\nu} $ & 4 \\
\hline
$S^{i}_\mu$ & 32/9 \\
$S^{4}_\mu$ & 4 \\
\hline
$B^{i4}_{\mu\nu}$ & 26/9 \\
$B^{ij}_{\mu\nu}$ & 34/9 \\
\hline
$J^i_{\mu\,j}$ & $3$ \\
$J^i_{\mu\,4}$ & $1+\tfrac{2}{9}\sqrt{77+4\sqrt{3}}=3.0358 \cdots$ \\
$J^4_{\mu\,4}$ & $1+\tfrac{4}{3}\sqrt{7/3}=3.0376 \cdots$ \\
\hline
$\chi^i$  & $110/27$ \\
\hline
$\chi^{i\bar{j}}_{\ 4}$ & $2$ \\
$\chi^{ij}_{\ \bar{k}}$ & $26/27+\tfrac{4}{3}\sqrt{2}=2.8486 \cdots$ \\
$\chi^{i4}_{\ \bar{4}}$ & $34/27+\tfrac{4}{3}\sqrt{2}=3.1449 \cdots$ \\
\hline
\end{tabular}
\end{center}
\caption{\small Conformal dimensions of the scaling operators
corresponding to the remaining
physical fields in the 5D supergravity multiplet at the $SU(3)$
point.}\label{tab:rest}
\end{table}

All the same remarks on the equations of motion in the previous section
apply in this case. It is consistent to take constant solutions for
$\rho,\alpha$. The same kind of asymptotic behavior for the metric and the
scalar $\psi$ as
$\mu z \to \infty$ is found, but with the conformal exponent being
now $b= 2^{1/2} 3^{1/3}$. The reason is the same: the scalar potential
\begin{equation*}
V_{SO(5)}(\psi) = -\frac{8}{ r^2}W(\psi)
\end{equation*}
has two critical points where $dV_{SO(5)}/d\psi=0$ (see fig.~\ref{fig:v}).

One critical point is at $\psi=0$ and it is a maximum of the potential.
It gives an AdS$_5$ space, with radius $r$,
coming from the compactification of type IIB on $S^5$.
On the 4D field theory side, it
corresponds to the $SO(6)$-invariant ${\cal N}=4$ superconformal vacuum.
The 10 dimensional geometry is the familiar one:
\begin{equation}\label{eq:rounds5}
ds^2 = r^2 \Bigl\{ \frac{1}{z^2} (dz^2 + \sum_{i=1}^4 dx_i^2)
+ d\alpha^2 + \sin^2\alpha \ d\Omega_{(4)} \Bigr\} \,.
\end{equation}
The anti-de Sitter space and the five sphere
have the same radius $r$. The five-form in the internal space is
\begin{equation}\label{eq:origfiveform}
F_5 = \frac{1}{r} \ d{\rm Vol}_{S^5}= r^4 \sin^4\alpha\
d\alpha\wedge d\Omega_{(4)} \,.
\end{equation}
The integral of $F_5$ over $S^5$ is quantized. In appropriate units, it gives
the number of D3-branes in the stack, whose near-horizon geometry is given by
\eqref{eq:rounds5}.

The other critical point is at $\psi=\psi_0$.
It breaks $SO(6) \to SO(5)$ and all the supersymmetries.
It was conjectured in \cite{Gunaydin:Gauged5dSUGRA} that it corresponds to the
compactification of type IIB on the inhomogeneously squashed five
sphere \cite{vanNieuwenhuizen:1985ri,Hull:GaugedSUGRAfromHigherDim}:
\begin{equation}\label{eq:10metric}
ds^2 = \tilde r^2\Bigl[ \frac{\sigma(\alpha)^2}{6z^2}(dz^2 +\sum_{i=1}^4
dx_i^2) +\frac{1}{12}(\sigma(\alpha)^2 d\alpha^2
+\sigma(\alpha)^{-2}\sin^2\alpha\ d\Omega_{(4)}^2)\Bigr]
\end{equation}
where
\begin{equation}
\sigma(\alpha)=(1-\frac{2}{3}\ \sin^2\alpha)^{1/4} \,;
\end{equation}
and the 5-form field strength
in the internal space is
\begin{equation}\label{eq:10fiveform}
F= \frac{1}{\tilde r} \sigma(\alpha)^{-5}\ d{\rm Vol}_{S^5} =
\tilde r^4 12^{-5/2} \sigma^{-8}\sin^4\alpha\ d\alpha\wedge d\Omega_{(4)} \,.
\end{equation}
The relation between the scale ${\tilde r}$, which characterizes the
$SO(5)$-invariant solution
\eqref{eq:10metric}, \eqref{eq:10fiveform}, and the scale $r$, which
characterizes the $SO(6)$-symmetric solution, is determined by noting that
the five-form flux, being quantized, must be preserved along the RG flow.
Equating the integrals of \eqref{eq:origfiveform} and \eqref{eq:10fiveform}
over $S^5$, we find
\begin{equation}\label{eq:rtildedef}
{\tilde r}= 2 \cdot 3^{3/8} r \,.
\end{equation}

Note, however, that this is not the same as the radius $r_{IR}= 2r/b$,
the radius of the infrared $AdS_5$ in the Einstein frame of the effective
5-dimensional theory. That is determined by normalizing the Einstein-Hilbert
term in the 5-dimensional effective action. One takes the 10-dimensional
Einstein-Hilbert term in the background metric \eqref{eq:10metric}, and
integrates it over $S^5$, and compares the result with the 5-dimensional
Einstein Hilbert term for the background metric
$ds^2=\frac{r_{IR}^2}{z^2}(dz^2+\sum_{i=1}^4dx_i^2)$ (but with
the 5-dimensional $M_{pl}$ determined from the original round metric
\eqref{eq:rounds5}). The integrals over $S^5$ which appear are identical (all
powers of the warp factor $\sigma(\alpha)$ disappear), and one has the relation
\begin{equation}
r_{IR}^3r^5= \frac{\tilde r^{10}}{(6\cdot12)^{5/2}}\ \frac{6}{\tilde r^2} \,.
\end{equation}
Plugging in \eqref{eq:rtildedef},
we recover $r_{IR}=\frac{2r}{2^{1/2}3^{1/3}}$.

This all sounds very promising, but we need to check the stability of this
supergravity solution before we can draw any conclusions.
As in the $SU(3)$ case, this requires that we compute the mass-squared of the
scalars about the new extremum of the potential. Since we have already
computed the potential for the scalars in the ${\bf 20}'$ of $SO(6)$, let us
compute  their masses first. As we saw, ${\bf 20}'={\bf 1}+{\bf 5}+{\bf 14}$.
The singlet is the field $\psi$ that we turned on. Clearly, its mass-squared is
positive at the new extremum (fig.~\ref{fig:v}). The ${\bf 5}$ are eaten by the
Higgs mechanism in the breaking of $SO(6)\to SO(5)$. They also have
positive mass-squared, as they are the longitudinal components of the massive
vector bosons. So we need to worry about the ${\bf 14}$. Unfortunately, that is
where trouble looms. The mass-squared of the ${\bf 14}$ can be
straightforwardly  computed from the potential
\eqref{eq:20pot} and one finds
\begin{equation}
m^2_{\bf 14} = -\frac{8}{3^{\frac{1}{3}} r^2}
< - \frac{2 \cdot 3^{\frac{2}{3}}}{r^2}\ \
\left(=-\frac{4}{r_{IR}^2}\right) \,;
\end{equation}
{\it i.e.}, they violate the stability bound. Put another way, if we
assumed that this solution corresponded to a CFT on the boundary, and
attempted to compute the conformal dimension of the operator corresponding to
the ${\bf 14}$, via
\begin{equation*}
\Delta(\Delta-4)=m^2 r_{IR}^2 \,,
\end{equation*}
we would find $\Delta_{\bf 14}$ to be {\it complex}.

{}From the supergravity side, there is nothing mysterious here. The
$SO(5)$ invariant supergravity simply does not provide a stable ground
state for
the theory. Fluctuations, {\it no matter how small}, cause it to decay.
And, by the same token, a classical solution with the asymptotics
\eqref{eq:asympSO(5)} is similarly destabilized.

With the benefit of hindsight, we should have \emph{expected} this instability
from the form of the perturbation \eqref{eq:pertso5}. Formally, it gives a
negative mass-squared to the scalars $X^1,\dots,X^5$ in the boundary
theory. The
naivest expectation for the resulting physics of the boundary theory is that it
runs off to infinity in field space and  has no stable vacuum. This expectation
appears to be confirmed by the supergravity analysis.

\section{Three Dimensional Conformal Field Theories}\label{sec:3D}

 The  $d=3$ ${\cal N}=8$ $SU(N_c)$ gauge theory
living on the world volume of $N_c$ coincident
M2 branes has an interacting infrared fixed point
that preserves all the supersymmetries \cite{Seiberg:Notes16Supercharges}.
The near horizon geometry of this BPS brane
configuration is $AdS_4 \times S^7$.
At large $N_c$, the CFT/AdS correspondence solves
the three dimensional SCFT through the mapping of its
generating functional to the one of
$D=4$ ${\cal N}=8$ $SO(8)$ gauged supergravity
\cite{deWit:SO(8)xSU(8)SUGRA} at the $SO(8)$ invariant background.
The local operators of the field theory are conveniently
mapped to the Kaluza-Klein spectrum of the
supergravity theory \cite{Aharony:AdSpxS11-p, Minwalla:AdS(4/7),
Entin:ChiralOps, Halyo:AdS4/7}.

The `massless' supermultiplet that includes the graviton
has 70 physical scalars arranged in the ${\bf 35_v} \oplus
{\bf 35_c}$ of $so(8)$ \cite{Casher:MassSpectrumS7,Gunaydin:S7Spectrum}.
They parametrize the coset space $E_{7(7)}/SU(8)$ via
a matrix $U$ in the 56 dimensional representation of $E_7$.
In the $SU(8)$ unitary gauge, it is given by
\begin{equation}
U(\phi) = {\rm exp}
\begin{pmatrix}
0 & \phi_{ijkl} \\ (\phi_{ijkl})^\star & 0
\end{pmatrix} \,,
\end{equation}
where $\phi_{ijkl}$ are 35 complex self-dual four-forms
($i,j,k,l=1, \dots, 8$ are $SO(8)$ indices).

At the supersymmetric $SO(8)$ invariant point $\phi_{ijkl}=0$,
these 70 scalars are tachyonic.
In the SCFT, they are mapped to relevant primary operators.
The ${\bf 35_v}$ scalars  (the real part of $\phi_{ijkl}$)
correspond to the $\Delta=1$ conformal primaries:
\begin{equation}\label{eq:deltaone}
{\cal O}_{(1)}^{ij}
= {\rm tr}(X^i X^j) - \frac{\delta^{ij}}{8}{\rm tr}(X^2) \,,
\end{equation}
and the ${\bf 35_c}$ pseudo-scalars
(the imaginary part of $\phi_{ijkl}$)  correspond to the $\Delta=2$
conformal primaries:
\begin{equation}\label{eq:deltatwo}
{\cal O}_{(2)}^{ij} = {\rm tr} (\lambda^i \lambda^j)
- \frac{\delta^{ij}}{8}{\rm tr}(\lambda\lambda) \,,
\end{equation}
where $X^i$ and $\lambda^i$ are the microscopic
(real) scalars
\footnote{They include the dualized vector field.}
and (Majorana) gauginos, in the irreps ${\bf 8_v}$
and ${\bf 8_c}$ of $SO(8)$ respectively, of the
$d=3$ ${\cal N}=8$ $SU(N_c)$ gauge theory.

Following the same philosophy as in the $d=4$ ${\cal N}=4$ SYM
case, in this section we
will discuss the relevant perturbations of the $d=3$ ${\cal N}=8$
fixed point via the addition of the
operators \eqref{eq:deltaone} and \eqref{eq:deltatwo} to the
supersymmetric Lagrangian.
The deformation of the theory through these relevant operators is driven
by the solution of the classical equations of motion,
subject to appropriate boundary conditions
for their associated supergravity modes $\phi_{ijkl}$.

As we have already learned in this paper,
the renormalization group trajectories of the relevant
operators will connect the ${\cal N}=8$ $SO(8)$
fixed point to other new fixed points, if
the supergravity potential admits additional stable
critical points, besides the $SO(8)$ invariant point at $\phi_{ijkl}=0$.

\subsection{\boldmath{$SU(3)$} invariant conformal field theories}

In \cite{Warner:NewExtremaScalarPot},
Warner performed an exhaustive study of
the  extrema of the ${\cal N}=8$ gauged $SO(8)$ supergravity
potential along $SU(3) \subset SO(8)$ invariant directions.
He found five additional critical points, displayed in table one
of \cite{Warner:NewExtremaScalarPot}.
At the level of eleven dimensional supergravity, they
correspond to compactifications $AdS_4 \times M_7$.
The four dimensional anti-de Sitter space ensures conformal
invariance for the field theory living at its boundary.
The compact seven dimensional manifold $M_7$
depends on the chosen critical point, such that for each one, it
breaks a  different number of supersymmetries and isometries.

There are three points where all the supersymmetries are
broken and the remaining isometries are $SO(7)^+$, $SO(7)^-$
and $SU(4)^-$. The stability of the $SO(7)^{\pm}$ invariant points
was checked in \cite{deWit:1984gs} and the answer was negative.
We do not know if the $SU(4)^-$ point is stable or not.

The other two points keep some unbroken supersymmetries,
which ensures vacuum stability
\cite{Breitenlohner:GaugedSUGRAstability,
Gibbons:StabilityGaugedSUGRA}.
Applying the CFT/AdS correspondence for these points,
they describe ${\cal N}=1$ $G_2$ and ${\cal N}=2$
$SU(3) \times U(1)$, invariant three-dimensional
superconformal field theories, where the correlation
functions are given by the supergravity partition
function evaluated at the appropriate $AdS_4 \times M_7$
background.

\subsection{The \boldmath{$SO(3) \times SO(3)$}
interacting fixed point}

 In \cite{Gibbons:StabilityGaugedSUGRA}, an $SO(3)$ invariant
critical point was found
for the $D=4$ ${\cal N}=5$ $SO(5)$ gauged supergravity theory.
The interesting thing about this particular point is that
it is a non-supersymmetric stable
anti-de Sitter background \cite{Boucher:PositiveEnergy},
which can be embedded in the ${\cal N}=8$ theory
\footnote{In \cite{Boucher:PositiveEnergy}
the stability is only proved for the ${\cal N}=5$ theory, but it was argued
that the result should generalize to
the full ${\cal N}=8$ theory.}
\cite{Warner:ScalarPotGaugedSUGRA}, where the unbroken isometries
are $SO(3)\times SO(3)$.
On the field theory side, it corresponds to
a new non-supersymmetric interacting fixed point
connected to the ${\cal N}=8$ through some direction
in the space of
$SO(3) \times SO(3)$ invariant renormalization group trajectories.

Decomposing the  ${\bf 35}_v \oplus {\bf 35}_c$ representation, first under
$SO(5)\times SO(3)$, and thence under $SO(3)\subset SO(5)$, one obtains:
\begin{equation}
\begin{split}
{\bf 35}_v+{\bf 35}_c&=2\bigl(({\bf 5},{\bf 1})+({\bf 10},{\bf 3})\bigr)\\
&=2\bigl(({\bf 3},{\bf 1})+2({\bf 1},{\bf 1})+3({\bf 3},
{\bf 3})+({\bf 1},{\bf 3})\bigr) \,.
\end{split}
\end{equation}
In particular, the $SO(3)$ singlets transform as a complex scalar in the
${\bf 5}$ of $SO(5)$. The truncation to the ${\cal N}=5$ theory consists in
setting to zero the scalars in the $({\bf 10},{\bf 3})$. Two complex
scalars in the
$({\bf 5},{\bf 1})$ can be turned on, breaking $SO(5)$ to $SO(3)$. In the
full theory,
this breaks $SO(8)$ to $SO(3)\times SO(3)$. In the process,  22 of the 70
scalars,
the $({\bf 3},{\bf 3})+2({\bf 1},{\bf 3})+2({\bf 3},{\bf 1})+({\bf 1},{\bf
1})$,
are eaten by the Higgs mechanism. The
remaining scalars are three singlets and five copies of the $({\bf 3},{\bf
3})$.

Let us denote the two complex scalars, which are turned on, by $\varphi_a$,
$a=1,2$. The  ${\cal N}=8$ supergravity potential reads
\cite{Warner:ScalarPotGaugedSUGRA}
\begin{equation}\label{eq:so3pot}
V (\boldsymbol{\varphi}) = - e^2 \left[ 2 + 4 f(|\boldsymbol{\varphi}|)
- \frac{1}{2}f^2(|\boldsymbol{\varphi}|) \left(|\boldsymbol{\varphi}|^4 -
|\varphi_1^2 + \varphi_2^2|^2 \right) \right]
\end{equation}
with $f(|\boldsymbol{\varphi}|) = 1/(1 - |\boldsymbol{\varphi}|^2)$
and $e^2$ the square of the $SO(8)$ gauge coupling, which is proportional
to the scalar curvature of the anti-de Sitter background
at the $SO(8)$ invariant point $\varphi_a =0$.

The kinetic terms for these scalars take the form
\cite{Warner:ScalarPotGaugedSUGRA}:
\begin{equation}\label{eq:KinSO(3)xSO(3)}
{\cal L}_{\mathrm{kin}}=f(|\boldsymbol{\varphi}|)^{3/2}
|\boldsymbol{\cal A}|^2 \,,
\end{equation}
where
\begin{equation}
{\cal A}^a_\mu=
\left(\delta^{ab}-\frac{1-f^{1/2}}{|\boldsymbol{\varphi}|^2}\varphi^a
\overline{\varphi^b}\right)\partial_\mu\varphi^b \,.
\end{equation}

The scalar potential \eqref{eq:so3pot} has an additional
extremum at
\begin{equation}\label{eq:so3point}
|\boldsymbol{\varphi}|^2 =4/5 \, , \quad \varphi_1^2 + \varphi_2^2 =0 \,.
\end{equation}

The methodology is the same as we learned in the four-dimensional
conformal field theories. The conformal structure exponent
$b$ at the $SO(3)\times SO(3)$ fixed point is different
from the one at the $SO(8)$ SCFT ($b\neq2$).
This exponent is given by the ratio of the
corresponding cosmological constants of the dual supergravity backgrounds
at these two critical points. For this case
\begin{equation}
\frac{b^2}{4} = \frac{\Lambda_{(SO(3)\times SO(3))}}
{\Lambda_{(SO(8))}} = \frac{7}{3} \,.
\end{equation}
\begin{figure}\centerline{\psfig{file=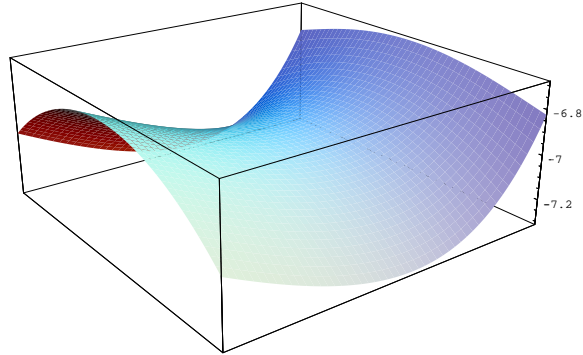,width=3.0in}}
\caption{The scalar potential \eqref{eq:so3pot}
near the $SO(3)\times SO(3)$ critical point.}
\label{fig:so3}
\end{figure}
In fig.~\ref{fig:so3} we plot the scalar potential along
two $SO(3) \times SO(3)$ singlet field directions, near the
critical point \eqref{eq:so3point}. As usual,
it is a saddle point. In CFT language,
the directions with negative (positive) mass-squared are mapped to relevant
(irrelevant) operators of the $SO(3) \times SO(3)$
fixed point. From the potential \eqref{eq:so3pot} and kinetic terms
\eqref{eq:KinSO(3)xSO(3)}, we can read off the masses of the scalars at this
point. From these, we derive the corresponding scaling dimensions using the
formula
\begin{equation}
\Delta (\Delta -3) = m^2 r_{IR}^2 \,.
\end{equation}
At the $SO(3)\times SO(3)$ CFT, we have
\begin{equation}
\Delta = \frac{3}{2} +
\sqrt{\frac{9}{4} + \frac{3}{7}m^2 r^2 }
\end{equation}
where $r = (\pi^2 N_c)^{1/6} l_p$ is the $AdS_4$ radius at the
supersymmetric $SO(8)$ invariant point and $m^2$ is the mass-squared
at the $SO(3)\times SO(3)$ point.

 The primary operator that drives the theory from the
$SO(8)$ SCFT to the $SO(3)\times SO(3)$ CFT is associated
to the common radial fluctuation $|\varphi_a| \to |\varphi_a|
+ \delta \rho$
and has scaling dimension
\begin{equation}
\Delta(\rho) = 8.146184\cdots
\end{equation}
Its partner, the common phase of the two complex scalars,
$\varphi_a\to\ex{i\alpha}\varphi_a$, is eaten by the Higgs mechanism.

The other two $SO(3)\times SO(3)$ singlet operators
continue to be tachyonic at this non-supersymmetric
conformal field theory. One is associated to the opposite
radial fluctuations $|\varphi_1| \to |\varphi_1| + \eta$,
 $|\varphi_2| \to |\varphi_2| - \eta$, and has scaling
dimension
\begin{equation}
\Delta(\eta) = 2.712887\cdots
\end{equation}
The other corresponds to the relative phase fluctuation
$\varphi_1 \to e^{i\beta} \varphi_1$,
$\varphi_2 \to e^{-i\beta}\varphi_2$
and has scaling dimension
\begin{equation}
\Delta(\beta) = 2.231925 \cdots
\end{equation}

\section{Six Dimensional Conformal Field Theories}\label{sec:6D}

The six dimensional $(2,0)$ SCFT
living on the world volume of $N$ coincident M5 branes
is described (at large N)
by 11D supergravity in the background AdS$_7 \times S^4$.
The R-symmetry groups is $Spin(5)$. As in 4 dimensions, one has chiral primary
fields, $\Phi_k$, consisting of $k^{th}$ order symmetrized traceless products
of the scalars in the  $N$ tensor multiplets (for $k=2,3,\dots$). These have
scaling dimension $\Delta_k=2k$. $\Phi_2$ is a relevant operator $(\Delta=4$)
in the $\bf{14}$ of $Spin(5)$. As in 4 dimensions, we can imagine perturbing
the theory by adding this relevant operator to the action, in such a way as to
break $Spin(5)$ to $Spin(4)$.

On the supergravity side,
this corresponds to giving a non-trivial boundary
conditions for an $SO(4)$-singlet scalar
in the $D=7$ ${\cal N}=4$ gauged supergravity multiplet.
The construction of the $SO(5)$ gauged $D=7$ ${\cal N}=4$
supergravity Lagrangian, up to two derivatives, was done in
\cite{Pernici:1984xx}. The supergravity potential turns out
to be equivalent to (\ref{eq:20pot}), but now with the matrices
$S$ valued in $SL(5,\BR)$ instead of $SL(6,\BR)$. As in the AdS$_5$ case, one
finds an additional  critical point with $SO(4)$ symmetry, corresponding
to a compactification on a squashed $S^4$.
But, as in the AdS$_5$ case,
this background is unstable \cite{Pernici:1985zw}. The scalars in the
supergravity multiplet in $\bf{14}$ break up into a $Spin(4)$ singlet, which
we are turning on, the $({\bf 2,2})$, which is higgsed, and the
$({\bf 3,3})$,
which violates the stability bound. Indeed, in analogy with what we
found in
\S\ref{sec:SO(5)}, one could have
\emph{expected} that this background would prove unstable. In the boundary
theory, the
$SO(4)$-invariant perturbation gives a positive mass-squared to one of the
scalars in the tensor multiplet, and a negative mass-squared to the other four.
So one expects this perturbation to cause the boundary theory to run off to
infinity in field space, with no stable vacuum.

The other possible four dimensional compact manifold with
$SO(4)$ isometries is $M_4 = Spin(4)/[U(1)\times U(1)] = S^2 \times S^2$.
In fact, recently it has been proposed to be
a non-supersymmetric stable vacuum of M-theory \cite{ Berkooz:1998qp}.
Here we perform an explicit check of that statement, computing part of
the Kaluza-Klein (KK) mass spectrum that results from that compactification.

For signature $(-+...+)$ and standard Ricci, the $D=11$
Einstein equations of motion are
\footnote{$A,B,...=0,...,10$ refer to eleven dimensional indices;
$\mu,\nu,...=1,...,4$ are internal indices of $S^2\times S^2$
and $a,b,...=0,...6$ correspond to indices in $AdS_7$.}
\begin{equation}
 R_{AB} = {1\over 6}  \left (F_{APQR}F_{B}{}^{PQR} -
            {1\over 12}g_{AB} F^2 \right) \,.
\label{eq:11d}
\end{equation}
We work in the Freund-Rubin ansatz:
\begin{equation}
\begin{split}
F_{\mu\nu\rho\sigma} &= e \, \epsilon_{\mu\nu\rho\sigma}
\\
F_{abcd} &= 0 \,,
\end{split}
\end{equation}
where $e$ is a constant proportional to the four-form
flux of field strength, created
by the presence of $N$ M5-branes, through the compact manifold $M_4$.
As it was noted in \cite{Berkooz:1998qp},
there is a particular solution with the topology
$AdS_7 \times S^2 \times S^2$.
We choose physical units where $r_{AdS_7}=1$.
Then, the equations (\ref{eq:11d}) determine $e=3{\sqrt 2}$
and the radius-squared of the two $S^2$ to be $r^2=1/12$.

There are three kinds of scalar fields. Two of them come from the
fluctuations of the eleven dimensional spacetime metric
\footnote{$\delta g^\mu{}_\mu$ ($\delta g_{(\mu\nu)}$)
means the trace(less) metric fluctuation.}
\begin{subequations}
\begin{align}
\delta g_{(\mu\nu)}(x^A) =
\sum_{I} \phi^{(I)}(x^a) Y^{I}_{(\mu\nu)}(x^\mu)
\label{eq:traceless}
\\
\delta g^{\mu}{}_{\mu}(x^A) = \sum_{I} \pi^{(I)}(x^a) Y^{I}(x^\mu) \,,
\label{eq:trace}
\end{align}
\label{eq:fluctua}
\end{subequations}
and the third from the eleven dimensional three-form potential
\begin{equation}
\delta A_{\mu\nu\rho}(x^A) =
\sum_{I} \chi^{(I)}(x^a) \epsilon_{\mu\nu\rho\sigma}
D_{\sigma} Y^{I}(x^\mu) \,.
\label{eq:3form}
\end{equation}

As usual, the internal space dependence is expanded
in a basis of tensor harmonics $Y$, $Y_{\mu}$ and $Y_{\mu\nu}$
of $M_4$,
with their coefficients being the seven dimensional scalar fields.
For the case of $M_4= S^2 \times S^2$,
these harmonics are a product, $Y = Y_1 Y_2$,
of the $S^2$ harmonics $Y_i^{(k_i)}$, $i=1,2$.
The index $I$ refer to the
$SO(4)\sim SU(2) \times SU(2)$ irreducible representations
$({\bf 2k_1 +1}, {\bf 2k_2 +1})$, for integer values of $k_1$ and $k_2$.
The masses of the scalar fields $\phi^{(I)}, \pi^{(I)}$ and $\chi^{(I)}$
are read from their linearized equations of motion obtained
from the variation of (\ref{eq:11d}), taking into account
expressions (\ref{eq:fluctua}) and (\ref{eq:3form}).
Since the scalar fluctuations (\ref{eq:trace}) change
the volume of $M_4$, this needs to be compensated by
a change in the three-form  (\ref{eq:3form}) in order
to keep the four-form field strength flux constant.
The consequence is that
the scalars $\pi^{(I)}$ and $\chi^{(I)}$ have mixed mass terms
and one should go to an appropiate basis which diagonalize them
\cite{vanNieuwenhuizen:1985iz}.

On the other hand, the traceless metric fluctuations (\ref{eq:traceless})
keep the volume of $M_4$ fixed. Then, it is consistent to
consider $F_{\mu\nu\rho\sigma}$ and $F^2$ constant.
Applying this kind of fluctuations to (\ref{eq:11d})
for the case of internal indices $A=\mu$ and $B=\nu$, one gets
\begin{equation}
\delta R_{\mu\nu} = 12 \,\delta g_{(\mu\nu)} \,.
\label{flu1}
\end{equation}

For a general metric fluctuation $\delta g_{AB}$, in a general background,
the fluctuation of the Ricci is (indices are raised and lowered
with the background metric)
\begin{equation}
\begin{split}
\delta R_{AB}= -{1\over 2} D^2
\delta g_{AB}
-{1\over 2}D_B D_A \delta g^P{}_P + {1\over 2} D_A D^P \delta g_{BP}
+ {1\over 2} D_B D^P \delta g_{AP}
\\
+ {1\over 2} R_A{}^P \delta g_{BP}
+{1\over 2} R_B{}^P \delta g_{AP} + R_{PABQ} \delta g^{PQ} \,.
\label{Riccifluc}
\end{split}
\end{equation}
Plugging this in (\ref{flu1}), it gives the equation (in de Donder gauge,
$D^\mu\delta g_{(\mu\nu)}=0$)
\begin{equation}\label{eq:fluct}
-{1\over 2}\square_{11} \delta g_{(\mu\nu)}
+ R_{\rho\mu\nu\sigma} \delta g^{(\rho\sigma)} =0 \,,
\end{equation}
with $\square_{11}$ the ordinary Laplacian for the $D=11$ metric $g_{AB}$.
Denoting the coordinate indices of the first $S^2$ by letters from the
beginning
of the Greek alphabet, and the coordinate indices of the second $S^2$ by
letters from the end of the Greek alphabet, the curvature tensor of
$S^2\times S^2$ is
\begin{equation}\label{eq:right}
\begin{split}
R_{\alpha\beta\gamma\delta}&=12(g_{\alpha\gamma}g_{\beta\delta}-
g_{\alpha\delta}g_{\beta\gamma})\\
R_{\phi\chi\psi\omega}&=12(g_{\phi\psi}g_{\chi\omega}-
g_{\phi\omega}g_{\chi\psi})
\end{split}
\end{equation}
with all of the mixed components vanishing. This is \emph{different} from all
of the other cases treated in \cite{vanNieuwenhuizen:1985iz}, where the
curvature tensor takes the form
\begin{equation}\label{eq:wrong}
R_{\kappa\lambda\mu\nu}=(g_{\kappa\mu}g_{\lambda\nu}-
g_{\kappa\nu}g_{\lambda\mu})
\end{equation}
In an earlier version of this paper, we naively applied the \emph{subsequent}
formul\ae\ of \cite{vanNieuwenhuizen:1985iz}, which implicitly assume the form
\eqref{eq:wrong}. In some cases, these gives results which agree with
\eqref{eq:right}, but in the crucial case of the breather mode, which we
discuss
below, they do not\footnote{We would like to acknowledge conversations with
Berkooz and Rey, which helped us track down the source of our previous
error.}.

Since we are dealing with a product space, there are two kind
of scalar modes from the traceless fluctuations
(\ref{eq:traceless}).
One kind correspond to the `jiggling' modes $\phi_j$, for
each of the two-spheres; {\it i.e.},
with $Y_{(\mu\nu)}$ being traceless for each of the $S^2$.
Their equation of motion are
\begin{equation}
(\square_{AdS_7} - m^2_j)\, \phi_j =0 \,,
\end{equation}
where the mass-squared is
\begin{equation}
m^2_j = -\square_4 + 24 \,.
\end{equation}
$\square_4$ are the eigenvalues of the
ordinary Laplace operator on the manifold $S^2 \times S^2$
acting on a symmetric and traceless two-tensor harmonic $Y_{\mu\nu}$:
\begin{equation}
\square_4 Y^{(k_1,k_2)}_{\mu\nu} =
-12\, [ k_1(k_1 +1) + k_2(k_2 +1) -2] Y^{(k_1,k_2)}_{\mu\nu}
\end{equation}
with $k_1 + k_2 \geq 2$ because it is a two-tensor harmonic.

The other kind of modes are the `breathing'
modes $\phi_b$, which are pure-trace
from the point of view of each of the two-spheres.
They come from the terms in the expansion (\ref{eq:traceless})
where $Y_{\alpha\beta}=Y g_{\alpha\beta}$
and $Y_{\psi\omega}=-Y g_{\psi\omega}$,
with $Y=Y^{(k_1)}_1Y^{(k_2)}_2$ being a product of \emph{scalar} harmonics on
each
$S^2$. The mass-squared of this breathing mode is
\begin{equation}
m^2_b = -\square_4 -24 \,,
\end{equation}
with now $\square_4 = 12\, [k_1(k_1 +1) + k_2(k_2 +1)]$,
and $k_1, k_2 = 0, 1,...$.
In particular, the lowest mode, corresponding to
$k_1=k_2=0$, has the squared mass $m_b^2= -24$,
which violates the stability bound $m^2 \geq -9$.
Therefore,
the non-supersymmetric background $AdS_7 \times S^2 \times S^2$ is unstable.

\bigskip

\section*{Acknowledgements}

We acknowledge valuable discussions with L.J. Boya,
R. Corrado and W. Fischler. We would also like to thank M. Berkooz and
S.~J.~Rey for discussions of their work, and for pointing out an error in a
previous version of this manuscript. F.Z. would like to thank CERN for its
support and hospitality, as well as useful discussions with L.
\'Alvarez-Gaum\'e, J.L.F. Barb\'on, M. Bremer, C. G\'omez and K. Skenderis,
during his visit in June.

\renewcommand{\baselinestretch}{1} \normalsize

\bibliography{ads}
\bibliographystyle{utphys}

\end{document}